\documentclass[prc,a4paper,showpacs,byrevtex]{revtex4}
\usepackage{graphicx}
\usepackage{dcolumn}
\usepackage{amsmath}
\usepackage{array}
\usepackage{bm}
\usepackage{amssymb}
\usepackage{amsfonts}

\begin{document}

\title{Eigenstates with the auxiliary field method}

\author{Claude Semay}
\thanks{F.R.S.-FNRS Senior Research Associate}
\email[E-mail: ]{claude.semay@umons.ac.be}
\affiliation{Service de Physique Nucl\'{e}aire et Subnucl\'{e}aire, Universit\'{e}
de Mons - UMONS, 20 Place du Parc,
7000 Mons, Belgium}
\author{Bernard Silvestre-Brac}
\email[E-mail: ]{silvestre@lpsc.in2p3.fr}
\affiliation{LPSC Universit\'{e} Joseph Fourier, Grenoble 1,
CNRS/IN2P3, Institut Polytechnique de Grenoble, 
Avenue des Martyrs 53, F-38026 Grenoble-Cedex, France}

\date{\today}

\begin{abstract}
The auxiliary field method is a powerful technique to obtain approximate closed-form energy formulas for eigenequations in quantum mechanics. Very good results can be obtained for Schr\"odinger and semirelativistic Hamiltonians with various potentials, even in the case of many-body problems. This method can also provide approximate eigenstates in terms of well known wavefunctions, for instance harmonic oscillator or hydrogen-like states, but with a characteristic size which depends on quantum numbers. In this paper, we consider two-body Schr\"odinger equations with linear, logarithmic and exponential potentials and show that analytical approximations of the corresponding eigenstates can be obtained with the auxiliary field method, with a very good accuracy in some cases. 
\end{abstract}

\maketitle

\section{Introduction}\label{intro}

A considerable amount of works have been already devoted to the computation of closed-form solutions of eigenequations in quantum mechanics, especially in bound states problems. Apart from its intrinsic mathematical interest, finding analytical formulas is always useful in physics, for example to obtain informations about the dependence of observables on the parameters of the Hamiltonian and on the quantum numbers. 

Many methods exist in the literature to find approximate analytical solutions of eigenequations for bound state problems: perturbation theory, Wentzel-Kramers-Brillouin (WKB) method, large-$N$ expansion, variational method, \textit{etc.} (see for instance Refs.~\cite{flu,dutt86} ). Recently, we have proposed a new technique, that we called the auxiliary field method (AFM), to compute bound states of a Schr\"odinger equation \cite{af}. This method is based on the introduction of auxiliary fields -- also known as einbein fields \cite{af1bis} -- into the original Hamiltonian and can lead to approximate closed-form results for various potentials \cite{af2,af3,hybri}. It has been extended to treat relativistic Hamiltonians \cite{afrela} and even many-body problems \cite{afnbody}. It is worth mentioning that we have shown in a previous work \cite{afmenv} that the AFM and the envelope theory \cite{hall0,hall1,hall2,hall3,hall4,hall5} are, to a large extent, two equivalent approaches. For instance, pure powers potential have been treated in detail in Refs.~\cite{af,hall3} and analytical results have been obtained for the Coulomb-plus-linear interaction in Refs.~\cite{af2,hall2}.

Since the principle of the AFM is to replace the studied Hamiltonian $H$ by a Hamiltonian $\tilde H$ for which analytical solutions are known, this method can also provide approximations for the genuine eigenfunctions of $H$ in terms of the eigenfunctions of $\tilde H$. The trial solutions are then characterized by a size which depends on the quantum numbers. The purpose of this paper is to test the quality of the trial wavefunctions given by the AFM by computing several observables, like mean values of powers of the operators position and square momentum. We consider two-body Schr\"odinger equations with three different potentials, relevant in particular for hadronic physics: linear, logarithmic, and exponential ones. We focus particularly on the case of the linear potential because analytical solutions are known for S-eigenstates. 

Our paper is organized as follows. The general principles of the AFM are recalled in Sec.~\ref{method}. The method is applied in Sec.~\ref{linear} to the case of the linear potential using two different approaches. Logarithmic and exponential potentials are studied in Secs.~\ref{ln} and \ref{exp} respectively. Finally we sum up our results in Sec.~\ref{conclu}. In order to be complete, a lot of mathematical expressions are given in this paper but, to avoid that technical details hamper the main text, they are relegated in appendices. Formulas to compute some observables are given in App.~\ref{A_obs}. Properties of Airy function, hydrogen-like system and harmonic oscillator are respectively reminded in App.~\ref{A_Ai}, \ref{A_Hy}, and \ref{A_OH}. The overlap of some dilated eigenstates are detailed in App.~\ref{A_overlap}. Finally, the inverse of a special function is also given in App.~\ref{A_invf}.

\section{The auxiliary field method}\label{method}

In order for the present paper to be self-contained, we recall here the main points of the AFM and refer the reader to Refs.~\cite{af,af2,af3,hybri,afrela,afnbody} for more details. The aim of this method is to find an analytical approximate solution of the eigenequation $H \left|\psi\right\rangle=E\left|\psi\right\rangle$, with
\begin{equation}\label{H}
H=T(\bm p^2)+V(r),
\end{equation}
where $V(r)$ is a central potential. Let us consider a new potential
\begin{equation}\label{Vtilde}
\tilde V(r,y)=y\, P(r)+V\left(I(y)\right)-y\, P\left(I(y)\right)
\end{equation}
for the Hamiltonian 
\begin{equation}\label{Htilde}
\tilde H(\hat \nu)=T(\bm p^2)+\tilde V(r,\hat \nu) ,
\end{equation}
which depends on the auxiliary field $\hat \nu$. Up to a dimensioned factor, $P(x)$ plays the role of another potential and $I(x)$ is the inverse of a function $K(x)$; explicitly, these functions are defined by 
\begin{equation}\label{defI}
I(x) =K^{-1}(x) \quad \textrm{and} \quad K(x)=\frac{V'(x)}{P'(x)},
\end{equation}
the prime denoting the derivative with respect to $x$. A proper elimination of this auxiliary field $\hat \nu$ by the following constraint $\left.\delta_{\hat \nu} H(\hat \nu)\right|_{\hat \nu=\hat\nu_0}=0$ leads to the solution $\hat\nu_0=K(r)$. The original Hamiltonian~(\ref{H}) can then be recovered since $\tilde V(r,\hat\nu_0)=V(r)$. 

The idea of the AFM is then to replace this operator $\hat\nu$ by a real parameter $\nu$. If the potential $P(r)$ is well chosen (for instance $P(r)=-1/r$ or $r^2$), the eigenequation  
\begin{equation}\label{HA}
H_A \left|\psi(\nu)\right\rangle=E_A(\nu)\left|\psi(\nu)\right\rangle
\end{equation}
with $H_A=T(\bm p^2)+\nu\, P(r)$ is analytically solvable for a nonrelativistic kinetic energy operator, as well as the one associated with the Hamiltonian 
\begin{equation}\label{htdef}
\tilde H(\nu)=T(\bm p^2)+\tilde V(r,\nu) .
\end{equation}
Let us remark that $\tilde V(r,\nu)$ is of the form $\nu P(r) + C(\nu)$ where $\nu$ and $C(\nu)$ are constants. Using~(\ref{HA}), an eigenvalue $E(\nu)$ of $\tilde H(\nu)$ is given by
\begin{equation}\label{Enu}
E(\nu)=E_A(\nu)+V\left(I(\nu)\right)-\nu\, P\left(I(\nu)\right).
\end{equation}
The approximate eigenvalues and eigenstates are eventually given by $E(\nu_0)$ and $\left|\psi(\nu_0)\right\rangle$ respectively, with $\nu_0$ such that $E(\nu_0)$ is extremal, \textit{i.e.}
\begin{equation}\label{enmin}
\left.\partial_\nu E(\nu)\right|_{\nu=\nu_0}=0.
\end{equation}
The value of $\nu_0$ depends on the quantum numbers of the state considered and should be denoted more precisely as $\nu_0(n,l)$. It is worth noting that if $P(r) = V (r)$, the method gives the exact results. Despite the fact that the auxiliary field is replaced by a real parameter, we keep the name ``auxiliary field method" used in previous studies \cite{morg99}. In these papers, the technique was applied to get rid of the square root operator appearing in semirelativistic Hamiltonians (see also Refs.~\cite{afrela,afnbody}).

Once $\nu_0$ is known, the constant $C(\nu_0)$ can be computed and is such that $\tilde V(r_0,\nu_0)=V(r_0)$ and $\tilde V'(r_0,\nu_0)=V'(r_0)$, where $r_0=I(\nu_0)$. It can be shown that \cite{af}
\begin{equation}\label{Pro}
\langle \psi(\nu_0) | P(r) | \psi(\nu_0) \rangle = P(r_0).
\end{equation}
This means that $r_0$ is a kind of ``average point" for the potential $P(r)$. Using this last relation with the definitions above, we get
\begin{equation}\label{Jrho}
\langle \psi(\nu_0) | J(\hat \nu_0) | \psi(\nu_0) \rangle = J(\nu_0)
\quad \textrm{with} \quad J(x)=P(I(x)).
\end{equation}
So, our method can actually be considered as a ``mean field approximation" with respect to a particular auxiliary field which is introduced to simplify the calculations: $\nu_0$ is the mean value of the operator $\hat \nu_0=K(r)$ through a function $J$ which can be quite simple. For example, $J(x)=x$ if $V(x)=P(x)^2/2+V_0$ where $V_0$ is a constant. 

Since the potential $\tilde V(r,\nu_0)$ is tangent to the potential $V(r)$ at $r=r_0$, the approximation $E(\nu_0)$ is an upper (lower) bound of the exact energy if $\tilde V(r,\nu_0) \ge V(r)$ ($\tilde V(r,\nu_0) \le V(r)$) for all values of $r$ \cite{afmenv}. Equivalently, a function $g(x)$ can be defined by $V(x)=g(P(x))$. It can then be shown that, if $g(x)$ is a concave (convex) function, the approximation $E(\nu_0)$ is an upper (lower) bound of the exact energy. This property has been demonstrated in the framework of the envelope theory \cite{hall0,hall1,hall2,hall3,hall4,hall5}, but can be applied as well to the AFM \cite{afmenv}. Several examples will be presented below. The knowledge of a lower and an upper bounds for an eigenstate is a first technique to estimate the accuracy of the AFM. It has also been shown that \cite{af}
\begin{equation}\label{eevv}
E(\nu_0) - \langle \psi(\nu_0) | H | \psi(\nu_0) \rangle = V(r_0) - \langle \psi(\nu_0) | V(r) | \psi(\nu_0) \rangle.
\end{equation}
The right-hand side of this equation is the difference between the value of potential $V$ computed at the average point $r_0$ and the average of this potential for the AFM state $| \psi(\nu_0) \rangle$ considered here as trial state. In some favorable cases (the trial state is a ground state for instance), $E \le \langle \psi(\nu_0) | H | \psi(\nu_0) \rangle$
and a bound on the error can be computed by 
\begin{equation}\label{relerr}
E(\nu_0) - E \ge V(r_0) - \langle \psi(\nu_0) | V(r) | \psi(\nu_0) \rangle.
\end{equation}
If the mean value of $V$ can be computed analytically, this constitutes a second procedure to estimate the accuracy of the AFM. Several examples of this calculation are presented in Ref.~\cite{af}. At last, the eigenstates of a Hamiltonian of type~(\ref{H}) can be solved numerically with an arbitrary precision. So, as a third possibility, a direct comparison with the AFM results is always possible. So, one can wonder why to use the AFM? The interest of this method is mainly to obtain analytical information about the whole spectra (dependence of eigenenergies on the parameters of the Hamiltonian and on the quantum numbers), without necessarily searching a very high accuracy.

The AFM is completely general and \emph{a priori} valid for any potential $V(r)$. Nevertheless, in order to get closed-form expressions, we must fulfill two conditions: first, to be able to determine the function $I(x)$ defined by (\ref{defI}) and, second, to be able to find $\nu_0$ defined by (\ref{enmin}) and to calculate the corresponding value $E(\nu_0)$ in an analytical way. Our previous works showed this can be managed for a lot of interesting physical problems and that very accurate approximate values can be obtained for the energy. 

Moreover, the state $\left|\psi(\nu_0)\right\rangle$, which is an eigenstate of $\tilde H(\nu_0)$, is an approximation of a genuine eigenstate of $H$. The shape of the corresponding wavefunction $\left\langle \bm r | \psi(\nu_0)\right\rangle$ depends on the quantum numbers via the parameter $\nu_0$. In the following sections, the quality of this approximation will be tested for the Schr\"odinger equation with different potentials by using the AFM with two different functions $P(r)=r^2$ and $-1/r$.

As it is shown in Ref.~\cite{af2}, the same energy formula $E(N)$ is obtained for the choice $P(r)=\textrm{sgn} (\lambda)r^\lambda$ with $\lambda > -2$, but with different forms for the global quantum number $N$. $N=2 n+l+3/2$ for $\lambda = 2$, and $N=n+l+1$ for $\lambda = -1$. For other values of $\lambda$, The form of $N$ is not exactly analytically known. 

\section{The linear potential}\label{linear}

Let us denote $m$ the mass of the particle for a one-body problem or the reduced mass for a two-body system.  When $T(\bm p^2)=\bm p^2/(2 m)$ and $V(r)=a r$ with $a$ positive, the Hamiltonian $H$ can be solved analytically for a vanishing angular momentum $l$ (see App.~\ref{A_Ai}). When $l\ne 0$, numerical results have been obtained with two different methods \cite{luch99,lagmesh}. In order to simplify the notation, we will denote $| n, l \rangle$ an eigenstate of $H$ and $| n \rangle=| n, 0 \rangle$, with $\langle \bm r | n \rangle$ given by~(\ref{psiAi}). The corresponding energies will be denoted $E_{n,l}=\langle n,l |H|n,l\rangle$ and $E_n=E_{n,0}=\langle n |H|n\rangle$. As we need a Hamiltonian $\tilde H$ with a central potential which is completely solvable to apply the AFM, we can only use in practice a hydrogen-like system ($P(r)=-1/r$) or a harmonic oscillator ($P(r)=r^2$). These two Hamiltonians will be used respectively in Sec.~\ref{analytHy} and \ref{analytOH}. 
As a linear potential seems closer to $r^2$ than $-1/r$, we can expect that the use of a harmonic oscillator to start the AFM will give better results. The scaling properties of the approximate AFM solutions and the exact solutions are the same \cite{af,af2}, so we can set $m=1/2$ and $a=1$ to lighten the notations and to match the results obtained in Ref.~\cite{af}. So, the following Hamiltonian is considered
\begin{equation}\label{Hlin}
H=\bm p^2 + r.
\end{equation}
We denote $\epsilon_{n,l}$ the approximate AFM energies which are given by \cite{af}
\begin{equation}\label{epsN}
\epsilon_{n,l}=\frac{3}{2^{2/3}}N^{2/3}.
\end{equation} 
Exact energies, given by (\ref{EAi}), reduces to $E_n= -\alpha_n$, where $\alpha_n$ is the $(n+1)^{\textrm{th}}$ zero of the Airy functions $\textrm{Ai}$. A simple approximation of $E_n$ can be obtained using the expansion (\ref{betan}) at the first order
\begin{equation}\label{Enex}
E_n= -\alpha_n \approx \left( \frac{3 \pi}{2}\right)^{2/3} \left( n + \frac{3}{4}\right)^{2/3} \approx 2.811 \left( n + \frac{3}{4}\right)^{2/3}.
\end{equation} 
It is worth noting that the linear potential is not only a toy model to test the AFM method. Effective theories of the quantum chromodynamics have proved that it is a good interaction to take into account the confinement of quarks or gluons in potential models of hadronic physics \cite{morg99,luch91}.

\subsection{AFM with $P(r)=-1/r$}\label{analytHy}

One can ask whether it is possible to obtain good approximations for the solutions of a Schr\"odinger equation with a linear potential by means of hydrogen-like eigenfunctions. This will be examined in this section. Using the results of Ref.~\cite{af} for $P(r)=-1/r$, we find $r_0=\sqrt{\nu_0}$ with $\nu_0=2^{2/3}(n+l+1)^{4/3}$. Exact eigenstates are then approximated by AFM eigenstates which are, in this case, hydrogen-like states (\ref{psiHy}) with 
\begin{equation}\label{etaHy}
\eta=\frac{\nu_0}{2}=\frac{(n+l+1)^{4/3}}{2^{1/3}}.
\end{equation} 
Such states are denoted $|\textrm{Hy};n,l\rangle$ and $|\textrm{Hy};n\rangle=|\textrm{Hy};n,0\rangle$. Using~(\ref{rkHy}) and results above, it can be shown that (\ref{Pro}) is satisfied, with $P(r_0)=-1/r_0=-2^{-1/3}(n+l+1)^{-2/3}$. It is also worth noting that (\ref{Jrho}) gives $\left\langle 1/\sqrt{\hat \nu} \right\rangle = 1/\sqrt{\nu_0}$. We denote $\epsilon_{n,l}^{\textrm{Hy}}$ and $\epsilon_n^{\textrm{Hy}}=\epsilon_{n,0}^{\textrm{Hy}}$ the approximated energies which are given by (\ref{epsN}) with $N=n+l+1$ \cite{af}. Since $\tilde V(r,\nu_0)-V(r)=-(r-r_0)^2/r \le 0$, $\epsilon_{n,l}^{\textrm{Hy}}$ are lower bounds of the exact energies. This can also be determined with the function $g$ defined in Sec.~\ref{method}. In this case, $g(y)=-1/y$ with $y<0$. The function $g''(y)=-2/y^3$ being positive, $g$ is convex as expected. 

The quantum  number dependence of the scaling parameter $\eta$ corrects partly the difference between the shapes of $\langle \bm r|n,l\rangle$ and $\langle \bm r|\textrm{Hy};n,l\rangle$. Consequently, $\langle\textrm{Hy};n,l |\textrm{Hy};n,l'\rangle=\delta_{ll'}$ because of the orthogonality of the spherical harmonics, but $\langle\textrm{Hy};n,l |\textrm{Hy};n',l\rangle\ne \delta_{nn'}$. Using the definition~(\ref{ovgen}), we find 
\begin{equation}\label{ovHy}
\langle\textrm{Hy};n,l |\textrm{Hy};n',l\rangle = F^{Hy}_{n,n',l}\left(\left(\frac{n'+l+1}{n+l+1}\right)^{4/3}\right)
\end{equation} 
with $F^{Hy}$ given by (\ref{FHy}). Table~\ref{tab:recHy} gives some values of $|\langle\textrm{Hy};n,l |\textrm{Hy};n',l\rangle|^2$. We can see that the overlap is not negligible for $n$ close to $n'$, but it decreases rapidly with $|n-n'|$. The situation improves when $l$ increases: $|\langle\textrm{Hy};0,l |\textrm{Hy};1,l\rangle|^2=0.43$, 0.29, 0.22, 0.18, 0.15, 0.13 for $l=0\to 5$.

\begin{table}[htb]
\protect\caption{Results for $P_{n,n',l}=|\langle\textrm{Hy};n,l |\textrm{Hy};n',l\rangle|^2$. Values for $l=0$ ($l=1$)
are given in the lower-left (upper-right) triangle of the Table. $P_{n,n',l}=P_{n',n,l}$ and $P_{n,n,l}=1$.}
\label{tab:recHy}
\begin{ruledtabular}
\begin{tabular}{rllll}
& $n=0$ & 1 & 2 & 3 \\
\hline
$n'=0$ & 1 & 0.29 & 0.028 & 0.0039 \\
1 & 0.43 & 1 & 0.36 & 0.036 \\
2 & 0.055 & 0.43 & 1 &  0.39 \\
3 & 0.0097 & 0.049 & 0.43 & 1
\end{tabular}
\end{ruledtabular}
\end{table}

\subsubsection{Results for $l=0$}

\begin{figure}[htb]
\includegraphics*[height=4cm]{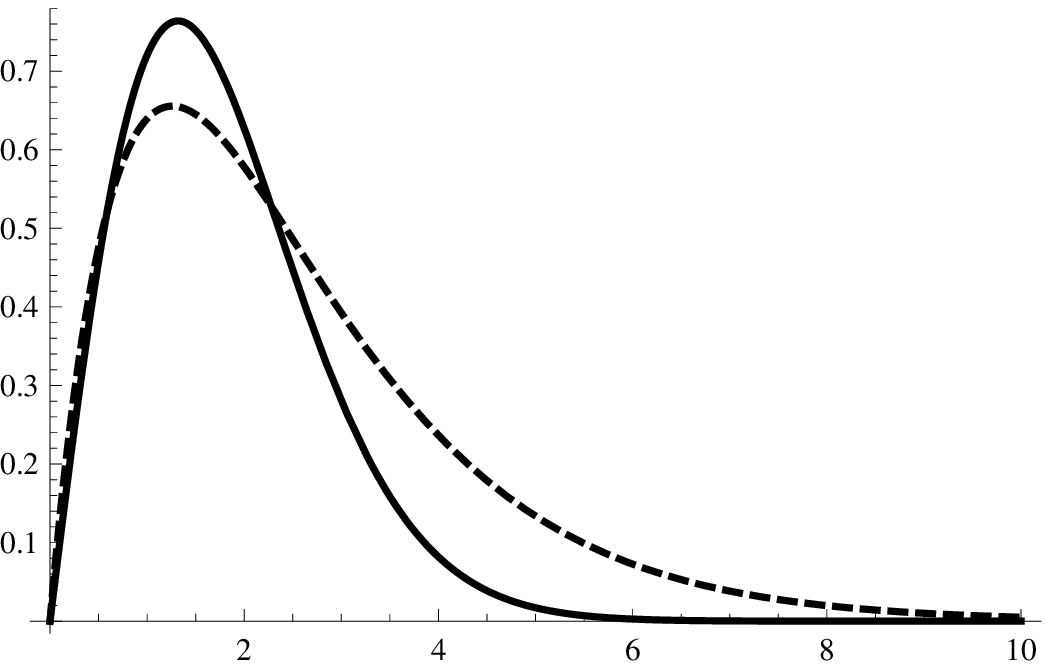}\quad
\includegraphics*[height=4cm]{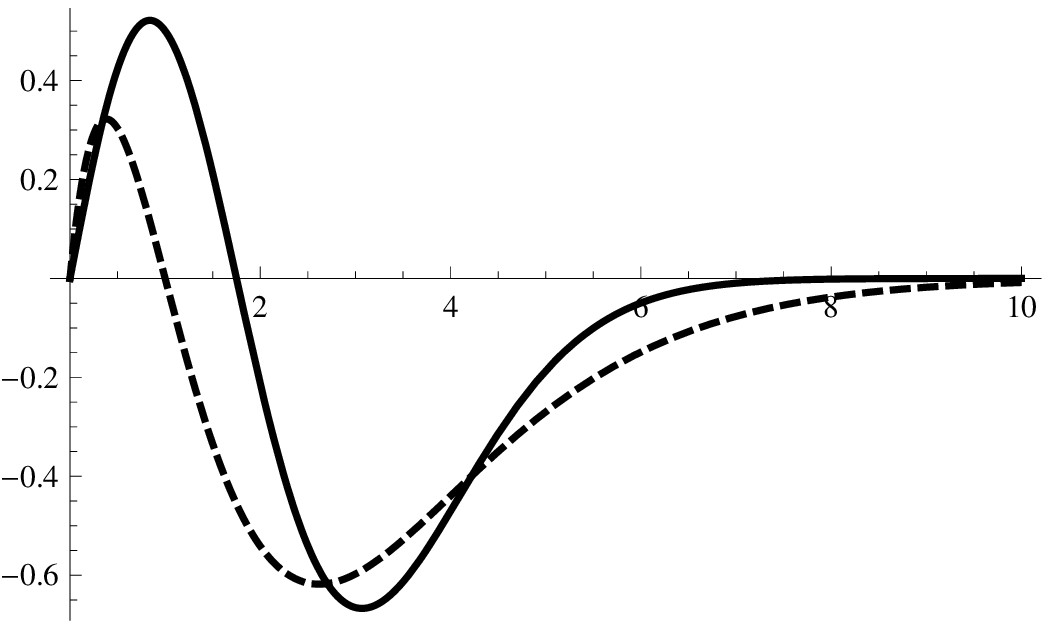}
\caption{Normalized wavefunctions $\langle \bm r|n\rangle$ (solid line) and $\langle \bm r|\textrm{Hy};n\rangle$ (dashed line) for $n=0$ (left) and 1 (right).}
\label{fig:Hy}
\end{figure}

It is interesting to compare $\epsilon_n^{\textrm{Hy}}$ with (\ref{Enex})
\begin{equation}\label{EepsHy}
\epsilon_n^{\textrm{Hy}}= \frac{3}{2^{2/3}}(n+1)^{2/3}\approx 1.890\, (n+1)^{2/3}. 
\end{equation}
The ratio $\epsilon_n^{\textrm{Hy}}/E_n$ is respectively equal to 0.808, 0.734, 0.712, for $n=0,1,2$ and tends rapidly toward the asymptotic value $3^{1/3}/\pi^{2/3}\approx 0.672$. As expected, these ratios are smaller than 1 since $\epsilon_n^{\textrm{Hy}}$ are lower bounds. Two wavefunctions are given in Fig.~\ref{fig:Hy}. We can see that the differences between exact $\langle \bm r|n\rangle$ and AFM $\langle \bm r|\textrm{Hy};n\rangle$ wavefunctions can be large. The overlap $|\langle n|\textrm{Hy};n\rangle|^2$ between these wavefunctions can be computed numerically with a high accuracy. We find respectively the values 0.934, 0.664, 0.298 for $n=0,1,2$, showing a rapid decreasing of the overlap. It is worth noting that, asymptotically, $\langle \bm r|n\rangle \propto \exp\left( -\frac{2}{3} \sqrt{2 m a}\,r^{3/2} \right)$ while $\langle \bm r|\textrm{Hy};n\rangle$ is characterized by an exponential tail. Nevertheless, if an observable is not too sensitive to the large $r$ behavior, this discrepancy will not spoil its mean value.

The various observables $|\psi_{n,0}(0)|^2$, $\langle r^k \rangle$ and $\langle p^k \rangle$ computed with the AFM states can be obtained using formulas~(\ref{psi0Hy}), (\ref{rkHy}) and (\ref{pkHy}) for $l=0$ with the parameter $\eta$ given by (\ref{etaHy}). Results are summed up in Table~\ref{tab:obsHy}. A direct comparison between the structure of exact and AFM observables can be obtained if we remind that the exact ones depend on $|\alpha_n|$ which can be well approximated by $\beta_n$ (see~(\ref{betan})). Let us look in detail only at the mean value $\langle r \rangle$. For the exact and AFM solutions, we have respectively
\begin{eqnarray}\label{rHy}
\label{rHy1} 
\langle n| r |n\rangle &=& \frac{2 |\alpha_n|}{3} \approx \left( \frac{2 \pi^2}{3}\right)^{1/3} \left( n + \frac{3}{4}\right)^{2/3} \approx 1.874 \left( n + \frac{3}{4}\right)^{2/3},\\
\label{rHy2} 
\langle \textrm{Hy};n| r |\textrm{Hy};n\rangle &=& \frac{3}{2^{2/3}}(n+1)^{2/3} \approx 1.890\, (n+1)^{2/3} . 
\end{eqnarray}
Some observables like $|\psi_{n,0}(0)|^2$ and $\langle p^4 \rangle$ are very badly reproduced. Others can be obtained with a quite reasonable accuracy. Despite the fact that exact and AFM wavefunctions differ strongly when $n$ increases, their sizes stay similar. 

Since $\langle\textrm{Hy};n|\bm p^2|\textrm{Hy};n\rangle$ and $\langle\textrm{Hy};n|r|\textrm{Hy};n\rangle$ are known, it is possible to compute analytically $\bar \epsilon_n^{\textrm{Hy}} = \langle\textrm{Hy};n|H|\textrm{Hy};n\rangle$. $|\textrm{Hy};n\rangle$, which is an exact eigenstate of the approximate Hamiltonian $\tilde H$, is then considered as a trial eigenstate for the genuine Hamiltonian $H$. The results, given in Table~\ref{tab:obsHy}, show that $\bar \epsilon_n^{\textrm{Hy}}$ is a better approximation than $\epsilon_n^{\textrm{Hy}}$. Because of the Ritz theorem, $\bar \epsilon_0^{\textrm{Hy}} \ge E_0$, but the variational character of other values $\bar \epsilon_n^{\textrm{Hy}}$ cannot be guaranteed, in contrast to the values $\epsilon_n^{\textrm{Hy}}$.

\begin{table}[hbt]
\protect\caption{Various observables $Q$ computed with the AFM and $P(r)=-1/r$. The ratios $\epsilon_n^{\textrm{Hy}}/E_n$ given in the last row can be compared with the ratios $\bar \epsilon_n^{\textrm{Hy}}/E_n$ given in the penultimate row.}
\label{tab:obsHy}
\begin{ruledtabular}
\begin{tabular}{cccccc}
 & $\langle\textrm{Hy};n|Q|\textrm{Hy};n\rangle$ & \multicolumn{4}{c}{$\langle\textrm{Hy};n|Q|\textrm{Hy};n\rangle/\langle n|Q|n\rangle$} \\
$\langle Q \rangle$ & ($2 m = a = 1$) & $n=0$ & $n=1$ & $n=2$ & $n\gg 1$ \\
\hline
$|\psi_{n,0}(0)|^2$ & $\frac{n+1}{2\pi}$ & 2 & 4 & 6 & $2(n+1)$ \\
$\langle r \rangle$ & $\frac{3(n+1)^{2/3}}{2^{2/3}}$ & 1.212 & 1.101 & 1.068 & $1.009 + \frac{0.168}{n}+ O\left(\frac{1}{n^2}\right)$ \\
$\langle r^2 \rangle$ & $\frac{5 n^2+10 n+6}{2^{1/3}(n+1)^{2/3}}$ & 1.633 & 1.178 & 1.080 & $0.942 + \frac{0.314}{n}+ O\left(\frac{1}{n^2}\right)$  \\
$\langle r^3 \rangle$ & $\frac{5 (7n^2+14 n+12)}{4}$ & 2.392 & 1.303 & 1.099 & $0.862 + \frac{0.431}{n}+ O\left(\frac{1}{n^2}\right)$  \\
$\langle r^4 \rangle$ & $\frac{3(21 n^4+84 n^3+161 n^2+154 n+60)}{2^{5/3}(n+1)^{4/3}}$ & 3.752 & 1.517 & 1.148 & $0.782 + \frac{0.522}{n}+ O\left(\frac{1}{n^2}\right)$  \\
$\langle p^2 \rangle$ & $\frac{(n+1)^{2/3}}{2^{2/3}}$ & 0.808 & 0.734 & 0.712 & $0.672 + \frac{0.112}{n}+ O\left(\frac{1}{n^2}\right)$  \\
$\langle p^4 \rangle$ & $\frac{(8 n+5)(n+1)^{4/3}}{2^{4/3}}$ & 1.815 & 3.890 & 5.916 & $2.009\, n + 1.926 + O\left(\frac{1}{n}\right)$  \\
$\langle H \rangle=\bar \epsilon_n^{\textrm{Hy}}$ & $\langle p^2 \rangle+\langle r \rangle$ & 1.078 & 0.978 & 0.949 & $0.896 + \frac{0.149}{n}+ O\left(\frac{1}{n^{4/3}}\right)$ \\
\hline & & 0.808 & 0.734 & 0.712 & $0.672 + \frac{0.112}{n}+ O\left(\frac{1}{n^{4/3}}\right)$ 
\end{tabular}
\end{ruledtabular}
\end{table}

\subsubsection{General results}

The behavior of observables computed with the AFM is similar for values of $l=0,1,2$. We do not expect strong deviations for larger values of $l$. This is illustrated with some typical results gathered in Table~\ref{tab:lne0Hy}. Observables are generally not very well reproduced, but it is expected since eigenstates for a linear potential are very different from eigenstates for a Coulomb potential. Actually, the agreement is not catastrophic, except for $|\psi_{n,0}(0)|^2$ and $\langle p^4 \rangle$ as mentioned above. It is even surprising that the AFM with $P(r)=-1/r$ could give energies and some observables for a linear potential with a quite reasonable accuracy. Moreover, lower bounds of the energies are obtained. This can be useful for some calculations (see Sec.~\ref{supp}).

\begin{table}[htb]
\protect\caption{Ratios between the AFM results (energies and $\langle r \rangle$) with $P(r)=-1/r$ and the exact results, for several quantum number sets $(n,l)$.}
\label{tab:lne0Hy}
\begin{ruledtabular}
\begin{tabular}{lllllll}
$l$ & $n=0$ &$n=1$ &$n=2$ &$n=3$ &$n=4$ &$n=5$ \\
\hline
\multicolumn{7}{l}{$\epsilon_{n,l}^{\textrm{Hy}}/E_{n,l}$ } \\
0 & 0.808 & 0.734 & 0.712 & 0.702 & 0.696 & 0.692 \\
1 & 0.893 & 0.805 & 0.767 & 0.746 & 0.733 & 0.724 \\
2 & 0.925 & 0.846 & 0.805 & 0.779 & 0.762 & 0.749 \\
\hline
\multicolumn{7}{l}{$\langle\textrm{Hy};n|r|\textrm{Hy};n\rangle/\langle n|r|n\rangle$} \\
0 & 1.212 & 1.101 & 1.068 & 1.053 & 1.043 & 1.037 \\
1 & 1.116 & 1.118 & 1.103 & 1.089 & 1.079 & 1.071 \\
2 & 1.080 & 1.110 & 1.110 & 1.104 & 1.096 & 1.089 \\
\end{tabular}
\end{ruledtabular}
\end{table}

\subsection{AFM with $P(r)=r^2$}\label{analytOH}

Since the quadratic potential is closer to the linear potential than a Coulomb one, one can expect better results with harmonic oscillator wavefunctions. This will be examined in this section. Using the results of Ref.~\cite{af} for $P(r)=r^2$, we find $r_0=1/(2\nu_0)$ with $\nu_0=2^{-4/3}(2 n+l+3/2)^{-2/3}$. Exact eigenstates are then approximated by AFM eigenstates which are, in this case, harmonic oscillator states (\ref{psiOH}) with 
\begin{equation}\label{lambHO}
\lambda=\nu_0^{1/4}=2^{-1/3}(2 n+l+3/2)^{-1/6}.
\end{equation} 
Such states are denoted $|\textrm{HO};n,l\rangle$ and $|\textrm{HO};n\rangle=|\textrm{HO};n,0\rangle$. Using~(\ref{rkOH}) and results above, it can be shown that (\ref{Pro}) is satisfied, with $P(r_0)=r_0^2=2^{2/3}(2 n+l+3/2)^{4/3}$. It is also worth noting that (\ref{Jrho}) gives $\left\langle 1/\hat \nu^2 \right\rangle = 1/\nu_0^2$. This is in agreement with (20) in Ref.~\cite{afrela} (an auxiliary field $\phi=1/\nu$ is used in this last reference). We denote $\epsilon_{n,l}^{\textrm{HO}}$ and $\epsilon_n^{\textrm{HO}}=\epsilon_{n,0}^{\textrm{HO}}$ the approximated energies which are given by (\ref{epsN}) with $N=2 n+l+3/2$ \cite{af}. Since $\tilde V(r,\nu_0)-V(r)=(r-r_0)^2/(2 r_0) \ge 0$, $\epsilon_{n,l}^{\textrm{HO}}$ are upper bounds of the exact energies. In this case, $g(y)=\sqrt{y}$ with $y>0$. The function $g''(y)=-1/(4 y^{3/2})$ being negative, $g$ is concave as expected.

The scaling parameter $\lambda$ depends on the quantum numbers. Consequently, $\langle\textrm{HO};n,l |\textrm{HO};n,l'\rangle=\delta_{ll'}$ because of the orthogonality of the spherical harmonics, but $\langle\textrm{HO};n,l |\textrm{HO};n',l\rangle\ne \delta_{nn'}$. Using the definition~(\ref{ovgen}), we find 
\begin{equation}\label{ovHO}
\langle\textrm{HO};n,l |\textrm{HO};n',l\rangle = F^{HO}_{n,n',l}\left(\left(\frac{4 n+2 l+3}{4 n'+2 l+3}\right)^{1/6}\right)
\end{equation} 
with $F^{HO}$ given by (\ref{FOH}). Table~\ref{tab:recOH} gives some values of $|\langle\textrm{HO};n,l |\textrm{HO};n',l\rangle|^2$. We can see that the overlap is always small and decreases rapidly with $|n-n'|$. The situation is even better when $l$ increases: $|\langle\textrm{HO};0,l |\textrm{HO};1,l\rangle|^2=0.029$, 0.023, 0.019, 0.017, 0.014, 0.013 for $l=0\to 5$.

\begin{table}[htb]
\protect\caption{Results for $P_{n,n',l}=|\langle\textrm{HO};n,l |\textrm{HO};n',l\rangle|^2$. Values for $l=0$ ($l=1$)
are given in the lower-left (upper-right) triangle of the Table. $P_{n,n',l}=P_{n',n,l}$ and $P_{n,n,l}=1$.}
\label{tab:recOH}
\begin{ruledtabular}
\begin{tabular}{rllll}
& $n=0$ & 1 & 2 & 3 \\
\hline
$n'=0$ & 1 & 0.023 & 0.0026 & 0.00039 \\
1 & 0.029 & 1 & 0.026 & 0.0027 \\
2 & 0.0036 & 0.027 & 1 &  0.026 \\
3 & 0.00064 & 0.0031 & 0.027 & 1
\end{tabular}
\end{ruledtabular}
\end{table}

\subsubsection{Results for $l=0$}

\begin{figure}[htb]
\includegraphics*[height=4cm]{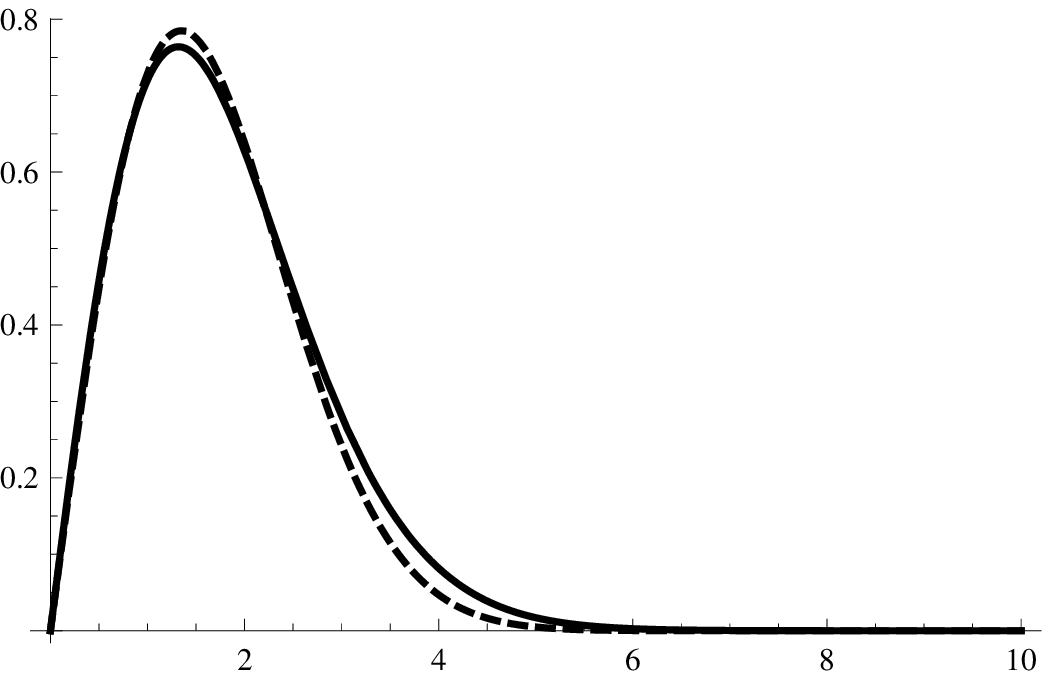}\quad
\includegraphics*[height=4cm]{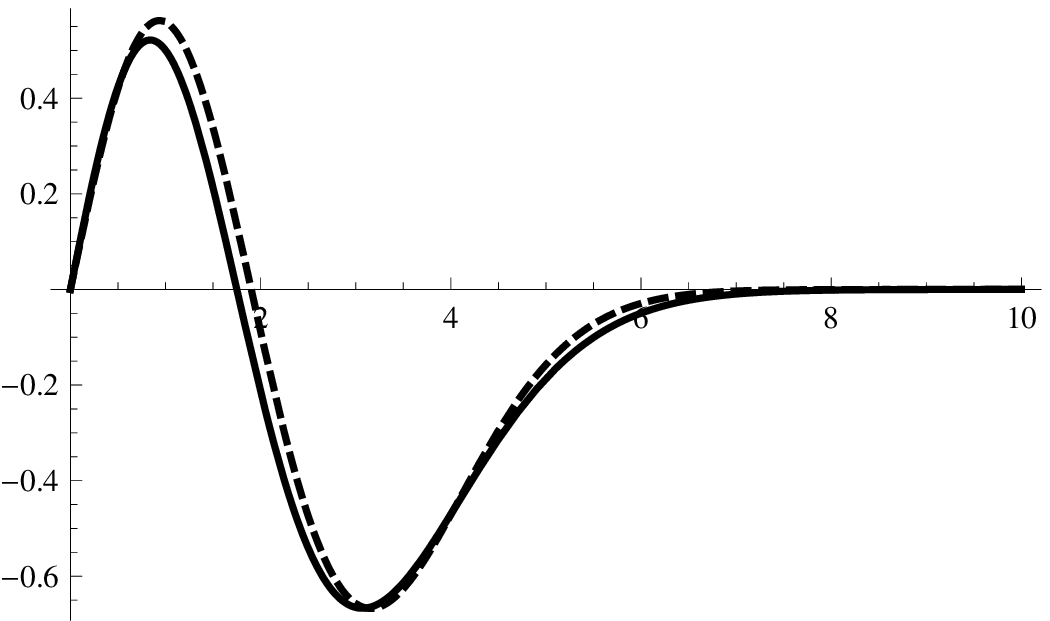}
\caption{Normalized wavefunctions $\langle \bm r|n\rangle$ (solid line) and $\langle \bm r|\textrm{HO};n\rangle$ (dashed line) for $n=0$ (left) and 1 (right).}
\label{fig:HO}
\end{figure}

Let us look at the energies $\epsilon_n^{\textrm{HO}}$ 
\begin{equation}\label{EespHO}
\epsilon_n^{\textrm{HO}}= 3 \left( n + \frac{3}{4}\right)^{2/3}. 
\end{equation}
By comparing with (\ref{Enex}), we can see immediately that the situation is more favorable than in the previous case. The ratio $\epsilon_n^{\textrm{HO}}/E_n$ is respectively equal to 1.059, 1.066, 1.067, for $n=0,1,2$. The asymptotic value $3^{1/3}2^{2/3}/\pi^{2/3}\approx 1.067$ is rapidly approached. As expected, these ratios are greater than 1 since $\epsilon_n^{\textrm{HO}}$ are upper bounds. Two wavefunctions are given in Fig.~\ref{fig:HO}. We can see that the differences between exact $\langle \bm r|n\rangle$ and AFM $\langle \bm r|\textrm{HO};n\rangle$ wavefunctions are quite small. The overlap $|\langle n|\textrm{HO};n\rangle|^2$ between these wavefunctions can be computed numerically with a high accuracy. We find respectively the values 0.997, 0.979, 0.951 for $n=0,1,2$. A value below 0.75 is reached for $n=6$ and below 0.25 for $n=14$. A wavefunction $\langle \bm r|\textrm{HO};n\rangle$ is characterized by an Gaussian tail, while we have asymptotically $\langle \bm r|n\rangle \propto \exp\left( -\frac{2}{3} \sqrt{2 m a}\,r^{3/2} \right)$. Again, if an observable is not too sensitive to the large $r$ behavior, this discrepancy will not spoil its mean value.

The various observables $|\psi_{n,0}(0)|^2$, $\langle r^k \rangle$ and $\langle p^k \rangle$ computed with the AFM states can be obtained using formulas~(\ref{psi0OH}), (\ref{rkOH}) and (\ref{pkOH}) for $l=0$ with the parameter $\lambda$ given by (\ref{lambHO}). Results are summed up in Table~\ref{tab:obsOH}. Again, a direct comparison between the structure of exact and AFM observables can be obtained by using $\beta_n$ (see~(\ref{betan})) instead of $|\alpha_n|$. Let us look in detail only at the mean value $\langle r^2 \rangle$. For the exact and AFM solutions, we have respectively
\begin{eqnarray}\label{rHO}
\label{rHO1} 
\langle n| r^2 |n\rangle &=& \frac{8|\alpha_n|^2}{15} \approx \frac{2^{5/3} 3^{1/3} \pi^{4/3}}{5} \left( n + \frac{3}{4}\right)^{4/3}\approx 4.214 \left( n + \frac{3}{4}\right)^{4/3},\\
\label{rHO2} 
\langle \textrm{HO};n| r^2 |\textrm{HO};n\rangle &=& 4 \left( n + \frac{3}{4}\right)^{4/3} . 
\end{eqnarray}
In contrast with the previous case, all observables are very well reproduced. This is also the case for $n\gg 1$, while the overlap $|\langle n|\textrm{HO};n\rangle|^2$ tends towards very small values in this limit. This is due to the fact that some observables are not very sensitive to the details of the wavefunctions and that the sizes of exact and AFM states stay similar.

In this case, it is also possible to compute analytically $\bar \epsilon_n^{\textrm{HO}} = \langle\textrm{HO};n|H|\textrm{HO};n\rangle$. The results, given in Table~\ref{tab:obsOH}, show that $\bar \epsilon_n^{\textrm{HO}}$ is a better approximation than $\epsilon_n^{\textrm{HO}}$. Because of the Ritz theorem, $\bar \epsilon_0^{\textrm{HO}} \ge E_0$, but the variational character of other values $\bar \epsilon_n^{\textrm{HO}}$ cannot be guaranteed, in contrast to the values $\epsilon_n^{\textrm{HO}}$.

\begin{table}[htb]
\protect\caption{Various observables $Q$ computed with the AFM and $P(r)=r^2$. The ratios $\epsilon_n^{\textrm{HO}}/E_n$ given in the last row can be compared with the ratios $\bar \epsilon_n^{\textrm{HO}}/E_n$ given in the penultimate row.}
\label{tab:obsOH}
\begin{ruledtabular}
\begin{tabular}{cccccc}
 & $\langle\textrm{HO};n|Q|\textrm{HO};n\rangle$ & \multicolumn{4}{c}{$\langle\textrm{HO};n|Q|\textrm{HO};n\rangle/\langle n|Q|n\rangle$} \\
$Q$ & ($2 m = a = 1$) & $n=0$ & $n=1$ & $n=2$ & $n\gg 1$ \\
\hline
$|\psi_{n,0}(0)|^2$ & $\frac{2 \Gamma(n+3/2)}{\pi^2 n! \sqrt{8 n+6}}$ & 0.921 & 0.905 & 0.902 & $0.900 + \frac{0.014}{n^2}+ O\left(\frac{1}{n^3}\right)$ \\
$\langle r \rangle$ & $\frac{4 (8 n+6)^{1/6} \Gamma(n+3/2)}{\pi n!}$ & 0.976 & 0.964 & 0.962 & $0.961 + \frac{0.011}{n^2}+ O\left(\frac{1}{n^{8/3}}\right)$ \\
$\langle r^2 \rangle$ & $4(n+3/4)^{4/3}$ & 0.935 & 0.946 & 0.948 & $0.949 - \frac{0.009}{n^2}+ O\left(\frac{1}{n^{7/3}}\right)$  \\
$\langle r^3 \rangle$ & $\frac{8 \sqrt{2} (4 n+3)^{3/2} \Gamma(n+3/2)}{3 \pi n!}$ & 0.881 & 0.934 & 0.941 & $0.946 - \frac{0.038}{n^2}+ O\left(\frac{1}{n^{7/3}}\right)$  \\
$\langle r^4 \rangle$ & $\frac{3}{4}(8 n+6)^{2/3}(8 n^2+12 n+5)$ & 0.819 & 0.918 & 0.934 & $0.946 - \frac{0.092}{n^2}+ O\left(\frac{1}{n^{7/3}}\right)$  \\
$\langle p^2 \rangle$ & $(n+3/4)^{2/3}$ & 1.059 & 1.066 & 1.067 & $1.067 - \frac{0.005}{n^2}+ O\left(\frac{1}{n^{7/3}}\right)$  \\
$\langle p^4 \rangle$ & $\frac{3(8 n^2+12 n+5)}{4(8 n+6)^{2/3}}$ & 1.039 & 0.966 & 0.956 &  $0.949 + \frac{0.050}{n^2}+ O\left(\frac{1}{n^{7/3}}\right)$ \\
$\langle H \rangle=\bar \epsilon_n^{\textrm{HO}}$ & $\langle p^2 \rangle+\langle r \rangle$ & 1.004 & 0.998 & 0.997 & $0.996 + \frac{0.005}{n^2}+ O\left(\frac{1}{n^{13/6}}\right)$ \\
\hline
 & & 1.059 & 1.066 & 1.067 & $1.067 - \frac{0.005}{n^2}+ O\left(\frac{1}{n^{7/3}}\right)$
\end{tabular}
\end{ruledtabular}
\end{table}

\subsubsection{General results}

As in the previous case, the behavior of observables computed with the AFM is similar for values of $l=0,1,2$. We do not expect strong deviations for larger values of $l$. Some typical results are presented in table~\ref{tab:lne0HO}. Agreement between AFM and exact results are very good, much better than for the previous case. This is expected, since eigenstates for a quadratic potential are closer to eigenstates for a linear potential than eigenstates for a Coulomb potential. The quantum number dependence of the scaling parameter $\lambda$ corrects, much better than $\eta$, the difference between the shapes of AFM and exact eigenstates. 

\begin{table}[htb]
\protect\caption{Ratios between the AFM results (energies and $\langle r \rangle$) with $P(r)=r^2$ and the exact results, for several quantum number sets $(n,l)$.}
\label{tab:lne0HO}
\begin{ruledtabular}
\begin{tabular}{lllllll}
$l$ & $n=0$ &$n=1$ &$n=2$ &$n=3$ &$n=4$ &$n=5$ \\
\hline
\multicolumn{7}{l}{$\epsilon_{n,l}^{\textrm{HO}}/E_{n,l}$ } \\
0 & 1.059 & 1.066 & 1.067 & 1.067 & 1.067 & 1.067 \\
1 & 1.036 & 1.055 & 1.060 & 1.063 & 1.064 & 1.065 \\
2 & 1.026 & 1.046 & 1.054 & 1.058 & 1.061 & 1.062 \\
\hline
\multicolumn{7}{l}{$\langle\textrm{HO};n|r|\textrm{HO};n\rangle/\langle n|r|n\rangle$} \\
0 & 0.976 & 0.964 & 0.962 & 0.962 & 0.961 & 0.961 \\
1 & 0.985 & 0.972 & 0.968 & 0.965 & 0.964 & 0.963 \\
2 & 0.990 & 0.978 & 0.972 & 0.969 & 0.967 & 0.966 \\
\end{tabular}
\end{ruledtabular}
\end{table}

\subsection{Supplementary topics}\label{supp}

For the linear potential, the energy formula obtained with the AFM is $\epsilon(N)=3(N/2)^{2/3}$, with $N=n+l+1$ for $P(r)=-1/r$ and with $N=2 n+l+3/2$ for $P(r)=r^2$. It has been shown that the accuracy of this energy formula can be greatly improved by changing the structure of $N$ \cite{af}. A very convenient form for the linear potential is (see (64) and (69) in Ref.~\cite{af})
\begin{equation}\label{Nimp}
N=\frac{\pi}{\sqrt{3}}\, n+l+\frac{\sqrt{3}\pi}{4},
\end{equation}
which gives
\begin{equation}\label{epsimp}
\epsilon_{n,l}= \left( \frac{3 \pi}{2}\right)^{2/3}\left( n+\frac{\sqrt{3}}{\pi}l+\frac{3}{4} \right)^{2/3}.
\end{equation}
This formula can be compared with (\ref{Enex}). The scaling parameters of the AFM eigenstates (see~(\ref{etaHy}) and (\ref{lambHO})), as well as the observables, depend explicitly on this global quantum number $N$. So, one can ask whether it is also possible to improve the accuracy of these observables by using the formula~(\ref{Nimp}) instead of the natural values of $N$. Unfortunately, this replacement spoils the values of some observables. We also tried to use a general form $N=A\, n+B\, l+C$ with $A$, $B$, $C$ as free parameters. In so doing, some observables can be improved, but, in the same time, others are dramatically spoilt. No unique set of values was found to improve globally the observables studied. Even in this case, no guarantee could be obtained about the behavior of others observables, and no systematic technique could be used to find the optimum parameters. So, if an improvement of the AFM energy can generally be obtained \cite{af,af2,af3,hybri,afrela,afnbody}, no such improvement seems possible for a set of observables.

The Eckart bound $B_E$ (see App.~\ref{A_obs}) is a lower bound on the overlap $|\langle \phi | 0 \rangle|^2$ between the exact ground state $| 0 \rangle$ and a trial state $| \phi \rangle$. In our case, it needs the computation of $\langle \phi |H| \phi \rangle$, with $| \phi \rangle=|\textrm{Hy};0\rangle$ or $|\textrm{HO};0\rangle$, and the knowledge of the exact energies for the ground and the first excited states. But all these values are known (see above). From a practical point of view, it is generally only possible to compute the less accurate bound $B'_E$ defined in App.~\ref{A_obs} since only lower and upper bounds on energies must be known. Such quantities can be computed with the AFM ($\epsilon_{n,l}^{\textrm{Hy}}$ are lower bounds and $\epsilon_{n,l}^{\textrm{HO}}$ are upper bounds), and the bound $B'_E$ can also be determined. Results, given in Table~\ref{tab:Eckart}, show that the bound $B'_E$, which is the easiest computable, cannot give relevant information in this case. On the contrary, the knowledge of the exact ground and first excited energies allows a good estimation of the overlap $|\langle \phi | 0 \rangle|^2$, without the necessity to compute the exact ground state $| 0 \rangle$.
\begin{table}[htb]
\protect\caption{Overlap $|\langle \phi | 0 \rangle|^2$, computed numerically, between a AFM state $| \phi \rangle$ and the exact ground state $| 0 \rangle$ compared with the lower bounds $B_E$ and $B'_E$ (see App.~\ref{A_obs}).}
\label{tab:Eckart}
\begin{ruledtabular}
\begin{tabular}{cccc}
$| \phi \rangle$ & $|\langle \phi | 0 \rangle|^2$ & $B_E$ & $B'_E$ \\
\hline
$|\textrm{Hy};0\rangle$ & 0.934 & 0.896 & 0.195 \\
$|\textrm{HO};0\rangle$ & 0.997 & 0.995 & 0.265 \\
\end{tabular}
\end{ruledtabular}
\end{table}

\section{The logarithmic potential}\label{ln}

The logarithmic potential, which has the particularity that its relative spectrum is independent of the mass
of the particle \cite{luch91,quig79}, can be used to simulate the confinement for heavy mesons. As the scaling properties of the approximate AFM solutions and the exact solutions are the same \cite{af,af2}, we can work with reduced variables. In order to match the results obtained in Ref.~\cite{af}, the following Hamiltonian is considered
\begin{equation}\label{Hln}
H=\frac{\bm p^2}{4}+\ln r
\end{equation}
Accurate numerical solutions have been obtained with two different methods \cite{luch99,lagmesh}. The notations of the previous section are reused to denote energies and eigenstates. 
Using the virial theorem, it can be shown that $\langle n,l|\bm p^2|n,l \rangle=2$ for any eigenstate. The logarithmic potential can be considered as the limiting case of a power potential when the exponent tends toward zero \cite{af,luch91,hall3}. So, in some sense, this interaction is ``midway" between parabolic and Coulomb potentials. It is then interesting to consider again both approximations given by $P(r)=-1/r$ and $r^2$. The approximated AFM energies are given by \cite{af}
\begin{equation}\label{lnepsN}
\epsilon_{n,l}=\ln \left( \sqrt{\frac{e}{2}}N \right).
\end{equation} 

\subsection{AFM with $P(r)=-1/r$}\label{lnHy}

Using the results of Ref.~\cite{af} for $P(r)=-1/r$, we find $r_0=\nu_0$ with $\nu_0=N/\sqrt{2}$ and $N=n+l+1$. AFM approximate eigenstates are hydrogen-like states (\ref{psiHy}) with 
\begin{equation}\label{lnetaHy}
\eta=2 \nu_0=\sqrt{2}N.
\end{equation} 
In this case, $g(y)=-\ln(-y)$ with $y<0$. The function $g''(y)=1/y^2$ being positive, $g$ is convex and $\epsilon_{n,l}^{\textrm{Hy}}$ are lower bounds of the exact energies. 

Mean values $\langle\textrm{Hy};n,l|r^2|\textrm{Hy};n,l\rangle$ and $\langle\textrm{Hy};n,l|p^2|\textrm{Hy};n,l\rangle$ are computed using (\ref{rkHy}) and (\ref{pkHy}) with (\ref{lnetaHy}). In particular we have $\langle\textrm{Hy};n,l|p^2|\textrm{Hy};n,l\rangle=2$. The AFM result is the exact one in this case. 

\subsection{AFM with $P(r)=r^2$}\label{lnHO}

Using the results of Ref.~\cite{af} for $P(r)=r^2$, we find $r_0=1/\sqrt{2 \nu_0}$ with $\nu_0=1/N^2$ and $N=2 n+l+3/2$. AFM approximate eigenstates are harmonic oscillator states (\ref{psiOH}) with 
\begin{equation}\label{lnlambHO}
\lambda=(4 \nu_0)^{1/4}=\sqrt{\frac{2}{N}}.
\end{equation} 
In this case, $g(y)=\ln(\sqrt{y})$ with $y>0$. The function $g''(y)=-1/(2 y^2)$ being negative, $g$ is concave and $\epsilon_{n,l}^{\textrm{HO}}$ are upper bounds of the exact energies. 

Mean values $\langle\textrm{HO};n,l|r^2|\textrm{HO};n,l\rangle$ and $\langle\textrm{HO};n,l|p^2|\textrm{HO};n,l\rangle$ are computed using (\ref{rkOH}) and (\ref{pkOH}) with (\ref{lnlambHO}). In particular we have $\langle\textrm{HO};n,l|p^2|\textrm{HO};n,l\rangle=2$. The AFM result is also the exact one in this case. 

\subsection{Results}\label{lnres}

Results concerning some observables derived from Hamiltonian~(\ref{Hln}) are gathered in Table~\ref{tab:lnres}. One can see that $\epsilon_{n,l}^{\textrm{Hy}}$ and $\epsilon_{n,l}^{\textrm{HO}}$ enclose the exact energy $E_{n,l}$, as expected. Lower and upper bounds are not very good, mainly for the ground state. They approach the exact values when $n$ or/and $l$ quantum numbers increase, and the relative error decreases below $10\%$ for the excited levels presented. The exact result is very close to the average between the lower and the upper bounds. It is worth noting that values of energies can be greatly improved (see Ref.~\cite{af}) but the variational character cannot be guaranteed anymore.

As mentioned, AFM reproduces the exact value of $\langle p^2 \rangle$ for both HO and Hy wave functions. For unknown reasons, AFM has sometimes very astonishing virtues like this one. The AFM $\langle r^2 \rangle$ values do not exhibit particular trends as function of the quantum numbers. They fluctuate around exact values but are maintained within satisfactory limits (less than $10\%$) for both HO and Hy wave functions.

The overlap between AFM and exact wave functions is rather good for $n=0$ but deteriorates rapidly with increasing values of $n$, the role of $l$ being much less important (a factor very important for the overlap is the number of nodes in the wave function). Here again HO and Hy wave functions give very similar results and behaviors.

It may seem strange that the quality of the energies increases sensitively while, in the same time, the quality of the overlap decreases drastically. The same conclusion is valid for other observables. This is due to the fact that the overlap is very dependent on the details of the wave function, but that the considered observables depends essentially on the  size of the wave function which is very similar for the exact and AFM wave functions. Contrary to the linear potential, the quality of the AFM results is quite similar whatever the choice of $P(r)$ indicating that the logarithmic potential, being midway the parabolic and the Coulomb potentials, can be approached equivalently with HO or Hy type wave functions. We have studied also other observables, as for the linear case, but the corresponding results leading essentially to the same conclusions, they are not presented here.

\begin{table}[htb]
\protect\caption{Some AFM results for the Hamiltonian~(\ref{Hln}): $R(E)=\epsilon_{n,l}^{\textrm{X}}/E_{n,l}$, $R(r^2)=\langle\textrm{X};n,l|r^2|\textrm{X};n,l\rangle/\langle n,l|r^2|n,l\rangle$, $R(p^2)=\langle\textrm{X};n,l|p^2|\textrm{X};n,l\rangle/\langle n,l|p^2|n,l\rangle$, and overlap $|\langle n,l|\textrm{X};n,l\rangle|^2$. For each set $(l,n)$, results are given for $P(r)=r^2$ (X=HO) in the first line and for $P(r)=-1/r$ (X=Hy) in the second line.}
\label{tab:lnres}
\begin{ruledtabular}
\begin{tabular}{lllllll}
$l$ & $n$ & X & $R(E)$ & $R(r^2)$ & $R(p^2)$ & $|\langle n,l|\textrm{X};n,l\rangle|^2$ \\
\hline 
0 & 0 & HO & 1.591 & 0.938 & 1 & 0.989 \\
  &   & Hy & 0.437 & 1.251 & 1 & 0.983 \\
  & 1 & HO & 1.218 & 1.051 & 1 & 0.766 \\
  &   & Hy & 0.733 & 0.900 & 1 & 0.771 \\
  & 2 & HO & 1.164 & 1.074 & 1 & 0.447 \\
  &   & Hy & 0.784 & 0.817 & 1 & 0.408 \\
1 & 0 & HO & 1.128 & 0.959 & 1 & 0.987 \\
  &   & Hy & 0.893 & 1.151 & 1 & 0.984 \\
  & 1 & HO & 1.137 & 1.014 & 1 & 0.792 \\
  &   & Hy & 0.859 & 1.002 & 1 & 0.835 \\
  & 2 & HO & 1.127 & 1.041 & 1 & 0.470 \\
  &   & Hy & 0.856 & 0.924 & 1 & 0.522 \\
2 & 0 & HO & 1.065 & 0.969 & 1 & 0.987 \\
  &   & Hy & 0.948 & 1.107 & 1 & 0.984 \\
  & 1 & HO & 1.097 & 0.998 & 1 & 0.815 \\
  &   & Hy & 0.909 & 1.039 & 1 & 0.866 \\
  & 2 & HO & 1.101 & 1.020 & 1 & 0.499 \\
  &   & Hy & 0.895 & 0.980 & 1 & 0.594 \\
\end{tabular}
\end{ruledtabular}
\end{table}

\section{The exponential potential}\label{exp}

The exponential potential is very different from the parabolic, linear, and logarithmic ones. It is not a confining interaction since continuum states are allowed. So, it is closer to the Coulomb interaction, with the difference that only a finite number of bound states exists. Let us note that the exponential potential can be used to simulate screening color effects in mesons \cite{gonz05}. In order to match the results obtained in Ref.~\cite{af3}, the following Hamiltonian written in reduced variables is considered
\begin{equation}\label{Hexp}
H=\bm p^2- k\, e^{-r}.
\end{equation}
Accurate numerical solutions have been obtained with two different methods \cite{luch99,lagmesh}. All bound states allowed have a negative energy. When the last excited states is characterized by an energy very close to zero, the corresponding wavefunction can have a very large extension. The notations of the previous section are reused to denote energies and eigenstates. Again, we consider both approximations given by $P(r)=-1/r$ and $r^2$. The approximated energies are given by
\begin{eqnarray}\label{expepsN}
\epsilon_{n,l}&=&-k\, e^{3 W_0(T)}\left( 1+\frac{3}{2} W_0(T)\right) \nonumber \\
&=& -k\,T^3 \left( \frac{1}{W_0(T)^3} + \frac{3}{2 W_0(T)^2} \right)
\quad \textrm{with} \quad T=-\frac{1}{3}\left( \frac{2 N^2}{k} \right)^{1/3},
\end{eqnarray} 
where $W_0$ is one branch of the Lambert function \cite{af3}. In this last paper, the formula is obtained using $P(r)=r$, but the calculations can be analytically managed for both for $P(r)=-1/r$ and $r^2$ to give again (\ref{expepsN}). The only technical difficulty is the inversion of the function $z = W(x) x^\alpha$ (see App.~\ref{A_invf}). Since negative values for the argument $T$ of $W_0$ must be in the range $[-1/e,0]$, only a limited set of values for $N$ are allowed. Good approximations for the critical values of $k$ can be obtained by imposing $\epsilon_{n,l}=0$ in (\ref{expepsN}) \cite{af3}.

\subsection{AFM with $P(r)=-1/r$}\label{expHy}

For $P(r)=-1/r$, we find $\nu_0=4 k u_0^2$ with $W_0(u_0)u_0^2=-N^2/(4 k)$ and $N=n+l+1$. AFM approximate eigenstates are hydrogen-like states (\ref{psiHy}) with 
\begin{equation}\label{expetaHy}
\eta=\frac{\nu_0}{2}=\frac{9 k}{2}\frac{T^3}{W_0(T)}.
\end{equation} 
In this case, $g(y)=-k e^{1/y}$ with $y<0$. Since the function $g''(y)=-k e^{1/y}(1/y^4+2/y^3)$ has not a constant sign for $y<0$, nothing can be said \emph{a priori} about the variational character of the eigenvalue (\ref{expepsN}) with $N=n+l+1$. But, it is possible to compute $\Delta V(r) = V(r,\nu_0)-V(r)= k [e^{-x}-r_0^2 e^{-r_0}/x + e^{-r_0}(r_0-1)]$ with $r_0=-3 W_0(T)$. A careful analysis shows that $\Delta V(r)$ is always negative if $r_0 \le 1$. This implies that $\epsilon_{n,l}^{\textrm{Hy}}$ are surely lower bounds of the exact energies if $n+l+1 \le \sqrt{k/(2 e)}$. Mean values $\langle\textrm{Hy};n,l|r^2|\textrm{Hy};n,l\rangle$ and $\langle\textrm{Hy};n,l|p^2|\textrm{Hy};n,l\rangle$ are then computed using (\ref{rkHy}) and (\ref{pkHy}) with (\ref{expetaHy}). 

\subsection{AFM with $P(r)=r^2$}\label{expHO}

For $P(r)=r^2$, we find $\nu_0=k/(2 u_0)$ with $W_0(u_0)u_0^{-1/4}=(2 N^2/k)^{1/4}$ and $N=2 n+l+3/2$. AFM approximate eigenstates are hydrogen oscillator states (\ref{psiOH}) with 
\begin{equation}\label{explambHO}
\lambda=\nu_0^{1/4}=\left( -\frac{k}{6}\frac{T^3}{W_0(T)^4} \right)^{1/4}.
\end{equation} 
In this case, $g(y)=-k e^{-\sqrt{y}}$ with $y>0$. The function $g''(y)=-k e^{-\sqrt{y}}(1/y+1/y^{3/2})/4$ being negative in its domain, $g$ is concave and $\epsilon_{n,l}^{\textrm{HO}}$ are upper bounds of the exact energies. Mean values $\langle\textrm{HO};n,l|r^2|\textrm{HO};n,l\rangle$ and $\langle\textrm{HO};n,l|p^2|\textrm{HO};n,l\rangle$ are then computed using (\ref{rkOH}) and (\ref{pkOH}) with (\ref{explambHO}). 

\subsection{Results}\label{expres}

Results concerning some observables derived from Hamiltonian~(\ref{Hexp}) are gathered in Table~\ref{tab:expres}. With $P(r)=r^2$, formula~(\ref{expepsN}) can give an imaginary or a positive value for the energy. When a negative energy is produced, it is an upper bound as expected but the agreement is poor. 

Better results are obtained for $P(r)=-1/r$. A negative value for the energy is produced for each allowed state, even if the agreement is sometimes poor. This happens for highly excited states with a near zero energy. A lower bound is alway obtained, though the variational character can only be guaranteed for the ground state in the cases $k=10$ and $k=20$. When the value of the energy is good, observables $\langle r^2 \rangle$ and $\langle p^2 \rangle$ are also quite good, as well as the overlap. Even if the results are not so accurate as in the previous cases, analytical informations about energies, observables, and the number of bound states can be obtained.

We have studied also other observables, as for the previous cases, but the corresponding results leading essentially to the same conclusions, they are not presented here. It is worth noting that values of energies can be greatly improved (see Ref.~\cite{af3}) but the variational character cannot be guaranteed anymore.

\begin{table}[htb]
\protect\caption{Some AFM results for the Hamiltonian~(\ref{Hexp}): $R(E)=\epsilon_{n,l}^{\textrm{X}}/E_{n,l}$, $R(r^2)=\langle\textrm{X};n,l|r^2|\textrm{X};n,l\rangle/\langle n,l|r^2|n,l\rangle$, $R(p^2)=\langle\textrm{X};n,l|p^2|\textrm{X};n,l\rangle/\langle n,l|p^2|n,l\rangle$, and overlap $|\langle n,l|\textrm{X};n,l\rangle|^2$. For each value of $k$, all allowed set $(l,n)$ are presented by increasing values of $E_{n,l}$, and, for each set, results are given for $P(r)=r^2$ (X=HO) in the first line and for $P(r)=-1/r$ (X=Hy) in the second line. The symbol ``-" indicates that the energy found is an imaginary or a positive number.}
\label{tab:expres}
\begin{ruledtabular}
\begin{tabular}{lllllllll}
$k$ & $l$ & $n$ & $E_{n,l}$ & X & $R(E)$ & $R(r^2)$ & $R(p^2)$ & $|\langle n,l|\textrm{X};n,l\rangle|^2$ \\
\hline 
5  & 0 & 0 & $-0.550$ & HO & 0.245 & 0.807 & 1.072 & 0.969 \\
   &   &   &          & Hy & 1.531 & 0.907 & 1.271 & 0.993 \\
   &   &   &          &    &       &       &       &       \\
10 & 0 & 0 & $-2.182$ & HO & 0.673 & 0.825 & 1.123 & 0.979 \\
   &   &   &          & Hy & 1.350 & 1.136 & 1.087 & 0.989 \\
   & 1 & 0 & $-0.334$ & HO & -     & -     & -     & -     \\
   &   &   &          & Hy & 1.370 & 0.738 & 1.346 & 0.973 \\
   & 0 & 1 & $-0.070$ & HO & -     & -     & -     & -     \\
   &   &   &          & Hy & 6.573 & 0.263 & 3.336 & 0.465 \\
   &   &   &          &    &       &       &       &       \\
20 & 0 & 0 & $-6.624$ & HO & 0.826 & 0.852 & 1.117 & 0.985 \\
   &   &   &          & Hy & 1.243 & 1.275 & 0.996 & 0.979 \\
   & 1 & 0 & $-2.715$ & HO & 0.647 & 0.876 & 1.068 & 0.975 \\
   &   &   &          & Hy & 1.242 & 1.005 & 1.117 & 0.992 \\
   & 0 & 1 & $-1.426$ & HO & -     & -     & -     & -     \\
   &   &   &          & Hy & 2.365 & 0.631 & 1.450 & 0.666 \\
   & 2 & 0 & $-0.431$ & HO & -     & -     & -     & -     \\
   &   &   &          & Hy & 1.251 & 0.773 & 1.271 & 0.974 \\
   & 1 & 1 & $-0.163$ & HO & -     & -     & -     & -     \\
   &   &   &          & Hy & 3.304 & 0.429 & 2.347 & 0.625 \\
   & 0 & 2 & $-0.009$ & HO & -     & -     & -     & -     \\
   &   &   &          & Hy & 62.1  & 0.066 & 11.90 &  0.043 \\
\end{tabular}
\end{ruledtabular}
\end{table}

\section{Concluding remarks}\label{conclu}

The auxiliary field method, which is strongly connected with the envelope theory \cite{hall0,hall1,hall2,hall3,hall4,hall5}, is a powerful tool to compute approximate closed-form energy formulas for eigenequations in quantum mechanics \cite{af}. The idea is to replace the potential studied $V(r)$ by a linear transformation of a well known potential $P(r)$, the coefficients of the transformation being state dependent. This method was already successfully applied for a great variety of problems \cite{af2,af3,hybri,afrela,afnbody}. It can also provide approximate analytical eigenstates of the Hamiltonian considered when a closed-form for the energy formula can be found. From a practical point of view, eigenstates for a central Hamiltonian can be approximated only by hydrogen-like ($P(r)=-1/r$) or harmonic oscillator ($P(r)=r^2$) wavefunctions which are properly scaled in order to match at best the studied eigenstates. 

In this paper, the quality of this approximation is tested for a Schr\"odinger equation with various potentials. For the linear potential, we found that very good results can be obtained for the wavefunction and a set of observables for the choice $P(r)=r^2$. The typical error on energy and other mean values is around 5-10\%. The error is quite independent of the quantum numbers and stay constant for large values of $n$. Quite good results can also be obtained for the logarithmic potential, which is a confining interaction as the linear one. With the choice $P(r)=-1/r$, analytical results have been found for the exponential potential but with an accuracy which is not as good as for the two previous cases. The main problem with this interaction is the existence of a finite number of bound states whose the most excited ones can have a near zero energy.

An eigenvalue equation can also be solved within the variational method by expanding trial states in terms of special basis states. The correct asymptotic tail can be well reproduced if the basis states are well chosen. With this method, a matrix representation of the Hamiltonian is obtained and the solutions are computed by diagonalizing this matrix: $M$ upper bounds of the energies are determined with the corresponding $M$ states, where $M$ is the order of the matrix. A very good accuracy is possible if $M$ is large enough. Even for $M=1$, the accuracy can be better than the one provided by the AFM \cite{afnbody}. However, if one is interested in closed-form results, the matrix elements must have an analytical expression and the number of computed states $M$ must be limited to 4. But even for $M=2$, the eigenvalues can have very complicated expression, not usable in practice. So, the variational method can only provide, at best, a very limited number of eigenvalues and eigenstates with an analytical form.

The WKB method is also a popular method to solve eigenvalue equations \cite{flu,griff}. In principle, it is only valid for high values of the radial quantum number $n$, but it can sometimes yield very good results for low-lying states \cite{brau00}. Wavefunctions are not necessary to compute some observables \cite{brau00} but they can be determined for arbitrary value of $n$ with this method. An advantage is that their asymptotic behavior can be correct but, unfortunately, the WKB method is mainly manageable for S-states. Indeed, for $l\ne 0$, the interaction $V(r)$ must be supplemented by the centrifugal potential, which complicates greatly the integrals to compute. Moreover, these wavefunctions are piecewise-defined whose different parts must be connected properly at the turning points. So they are not very practical to use. 

Within the auxiliary field method, if the problem studied is analytically manageable, a state can be determined with the same calculatory effort for any set of quantum numbers ($n,l$). So this method is very useful if one is interested in obtaining analytical information about the whole spectra, wavefunctions, and observables of a Hamiltonian without necessarily searching a very high accuracy. Moreover, a AFM state can be used as a trial state to recompute the energy, like in the variational method. But, in this case, an upper bound can be guaranteed only for the ground state.  

It is shown in this work that the selection of the potential $P(r)$ is crucial to obtain good results. For a linear potential, $P(r)=r^2$ is the best choice. But $P(r)=-1/r$ provides much better results for an exponential potential, while the choice of $P(r)$ is not so crucial for the logarithmic interaction. Other kind of potentials could be considered within the AFM. It could also be interesting to look at Hamiltonians with relativistic kinematics. 

\section*{Acknowledgments}

C. Semay thanks F. Buisseret and F. Brau for useful discussions. He also thanks the F.R.S.-FNRS for financial support. 

\appendix

\section{Observables}\label{A_obs}

In this appendix, we recall general properties for observables that are studied in this work. A given state is denoted $|E\rangle$ or $|n, l\rangle$ where $n$ is the radial quantum number and $l$ the orbital angular momentum. In order to simplify the notations, the mean value $\langle n, l|Q|n, l \rangle$ of an observable $Q$ is sometimes simply denoted $\langle Q \rangle$, and the abbreviation $| n \rangle=| n, 0 \rangle$ is also used. 

\textit{Overlap with the ground state}

Let $H$ be a Hamiltonian and $|E_i\rangle$ one of its eigenstate with energy $E_i$. The Eckart bound $B_E$ \cite{ecka30} gives a lower bound on the overlap between the state $|E_0\rangle$ and a trial state $| \phi \rangle$. This bound necessitates the knowledge of the exact values of $E_0$ and $E_1$. But it is possible to replace $B_E$ by a new bound $B'_E$ less accurate by using upper $E_i^U$ and lower $E_i^L$ bounds of a given level $E_i$. This gives
\begin{equation}\label{BE}
|\langle \phi | E_0 \rangle|^2 \ge B_E = \frac{E_1 - \langle \phi | H | \phi \rangle}{E_1 - E_0} \ge
B'_E = \frac{E_1^L - \langle \phi | H | \phi \rangle}{E_1^U - E_0^L}.
\end{equation}  

\textit{Mean values of even powers of the momentum}

Let us consider a Hamiltonian of type
\begin{equation}\label{HNR}
H = \frac{\bm p^2}{2 m}+ V(r),
\end{equation}
where $V(r)$ is a central potential with $r=|\bm r|$. Using the Virial theorem \cite{luch91} and the hermiticity of $H$ in the computation of 
$\langle H^2 \rangle$, one finds
\begin{eqnarray}\label{p2p4}
\label{p2} 
&& \langle \bm p^2 \rangle = 2 m \left( E - \langle V \rangle \right) = m \langle r\, V' \rangle, \\
\label{p4} 
&& \langle \bm p^4 \rangle = 4 m^2 \left( E^2 - 2 E \langle V \rangle + \langle V^2 \rangle \right), 
\end{eqnarray}
with the notation $Q'=dQ/dr$ and $E=\langle H \rangle$. Simplifications occur if the potential is a homogeneous function of degree $\lambda$ ($V(\beta r) = \beta^\lambda V(r)$), in which case $r\, V' = \lambda\, V$, so that these equations reduce to
\begin{eqnarray}\label{p2p4bis}
\label{p2b} 
&& \langle \bm p^2 \rangle = \frac{2 m \lambda}{\lambda+2} E \quad \textrm{and} \quad 
\langle V \rangle = \frac{2 }{\lambda+2} E, \\
\label{p4b} 
&& \langle \bm p^4 \rangle = 4 m^2 \left( \frac{\lambda-2}{\lambda+2} E^2 + \langle V^2 \rangle \right). 
\end{eqnarray}
This is in particular the situation for a power-law potential.

\textit{Wavefunction at the origin}

The square of the modulus of the wavefunction $\psi_{n,l}(\bm r)=\langle \bm r |n,l \rangle$ at the origin for a $l=0$ state (for $l \neq 0$, it vanishes) can be computed with a formula given in Ref.~\cite{luch91}:
\begin{equation}
| \psi_{n,0}(0) |^2 =\frac{m}{2\pi} \langle V' \rangle.
\end{equation}

\textit{Generalized virial theorem}

If $A$ is an arbitrary operator, it follows that
\begin{equation}\label{meanHA}
\langle [H,A] \rangle=0,
\end{equation}
due to the hermiticity of $H$. If $p_r$ is the radial momentum (with $[r,p_r]=i$ and $\hbar=1$) and $f(r)$ an arbitrary function, the computation of (\ref{meanHA}) with $A=p_r f(r)$ yields to the following relation, called the generalized virial theorem,
\begin{equation}\label{vir1}
\left\langle 2 E f'(r)-V'(r) f(r) - 2 V(r) f'(r) 
+ \frac{f'''(r)}{4 m} - \frac{l(l+1)}{m r} \left( \frac{f(r)}{r} \right)' \right\rangle =0,
\end{equation}
One recovers the usual virial theorem (see (\ref{p2})) for the special choice $f(r)=r$. But new interesting relations can be obtained for other choices. In particular, if $f(r)=r^{s+1}$ the previous equation becomes
\begin{equation}\label{vir3}
2 (s+1) E \langle r^s \rangle - 2 (s+1) \langle r^s V(r) \rangle - \langle r^{s+1} V'(r) \rangle 
+ \frac{s}{4 m} \left( s^2-1-4 l (l+1) \right) \langle r^{s-2} \rangle = 0.
\end{equation}
In addition, if $V(r)=\textrm{sgn}(\lambda) a r^\lambda$ for $\lambda \ne 0$, (\ref{vir3}) reduces to
\begin{equation}\label{vir5}
2 (s+1) E \langle r^s \rangle - \textrm{sgn}(\lambda) a (2 s+\lambda+2) \langle r^{\lambda+s} \rangle  
+ \frac{s}{4 m} \left( s^2-1-4 l (l+1) \right) \langle r^{s-2} \rangle = 0.
\end{equation}
This recurrence relation is particularly useful to compute $\langle r^k \rangle$ mean values for potentials with integer power, choosing $s$ as an integer.

\section{Exact solutions for the linear potential}\label{A_Ai}

The eigensolutions with $l=0$ of the Hamiltonian
\begin{equation}\label{HAi}
H = \frac{\bm p^2}{2 m}+  a \, r
\end{equation}
are analytically known in terms of the Airy functions $\textrm{Ai}$ \cite{abra}. The eigenenergies can be written in terms of the (negative) zeros $\alpha_n$ of this function, namely ($\hbar=1$)
\begin{equation}\label{EAi}
E = -\left( \frac{a^2}{2 m} \right)^{1/3} \alpha_n,
\end{equation}
and the normalized eigenvectors $\psi_{n0}(\bm r)=\langle \bm r |n\rangle$ are given by
\begin{equation}\label{psiAi}
\psi_{n0}(\bm r)=\frac{\left( 2 m a \right)^{1/6}}{\sqrt{4 \pi} \left| \textrm{Ai}'(\alpha_n) \right| r}
\textrm{Ai} \left( \left( 2 m a \right)^{1/3} r + \alpha_n \right).
\end{equation}
An approximate form for $\alpha_n$ is given by \cite{abra} 
\begin{equation}\label{betan}
\alpha_n = -\beta_n \left( 1+ \frac{5}{48} \beta_n^{-3} - \frac{5}{36} \beta_n^{-6} + O(\beta_n^{-9}) \right) \quad \textrm{with} \quad \beta_n =\left[ \frac{3 \pi}{2} \left( n + \frac{3}{4}\right) \right]^{2/3},
\end{equation}
the series converging very rapidly with $n$. At the origin, the square modulus of the wavefunction reduces to
\begin{equation}\label{psi0Ai}
|\psi_{n,0}(0)|^2=\frac{m a}{2 \pi}.
\end{equation}
Remarkably this quantity does not depend on the radial quantum number. This property is specific to the linear potential.

Mean values $\langle r^k \rangle$ can be computed by performing directly the integrals or by using (\ref{vir5}). One obtains
\begin{eqnarray}\label{rkAi}
\langle n|r |n\rangle &=& \frac{2 |\alpha_n|}{3(2 m  a)^{1/3}}, \nonumber \\
\langle n|r^2 |n\rangle &=& \frac{8 |\alpha_n|^2}{15(2 m  a)^{2/3}}, \nonumber \\
\langle n|r^3 |n\rangle &=& \frac{16 |\alpha_n|^3+15}{35(2 m  a)}, \nonumber \\
\langle n|r^4 |n\rangle &=& 16 \frac{8 |\alpha_n|^4+25 |\alpha_n|}{315(2 m  a)^{4/3}}.
\end{eqnarray}
Using (\ref{p2b}) and (\ref{p4b}), it is easy to show that 
\begin{eqnarray}\label{pkAi}
\langle n|p^2 |n\rangle &=& (2 m  a)^{2/3}\frac{|\alpha_n|}{3}, \nonumber \\
\langle n|p^4 |n\rangle &=& (2 m  a)^{4/3}\frac{|\alpha_n|^2}{5}.
\end{eqnarray}

\section{Hydrogen-like system}\label{A_Hy}

The eigenenergies of a hydrogen-like Hamiltonian \cite{flu,brau99}
\begin{equation}\label{HHy}
H = \frac{\bm p^2}{2 m} -  \frac{\nu}{r},
\end{equation}
are given by ($\hbar=1$)
\begin{equation}\label{EHy}
E = -\frac{m \nu^2}{2 N_{Hy}^2}  \quad \textrm{with} \quad N_{Hy}= n+l+1,
\end{equation}
and the normalized eigenvectors by
\begin{equation}\label{psiHy}
\psi_{nl\mu}(\bm r)=\left(2 \gamma_{nl}\right)^{3/2}\sqrt{\frac{n!}{2(n+l+1)(n+2 l+1)!}}\left( 2 \gamma_{nl} r \right)^l e^{-\gamma_{nl} r} L_{n}^{2 l+1}(2 \gamma_{nl} r) Y^l_\mu(\hat{\bm{r}}),
\end{equation}
with $\gamma_{nl}= \eta/(n+l+1)$ and $\eta= m \nu$. $L_{\alpha}^{\beta}$ is a Laguerre polynomial and $Y^l_\mu$ a spherical harmonic. At the origin, the S-states are such that 
\begin{equation}\label{psi0Hy}
|\psi_{n,0}(0)|^2=\frac{\eta^3}{\pi (n+1)^3}.
\end{equation}
Mean values $\langle r^k \rangle$ can be computed by performing directly the integrals or by using (\ref{vir5}). One obtains
\begin{equation}\label{rkHygen}
\langle r^k \rangle = \frac{(n+l+1)^{k-1}}{2(2\eta)^k}\frac{(n+2 l+1)!}{n!}\sum_{p,q=0}^n (-1)^{p+q}C_n^p C_n^q \frac{(p+q+k+2 l+2)!}{(p+2 l+1)!(q+2 l+1)!},
\end{equation}
where $C_\alpha^\beta$ is the usual binomial coefficient. With $L=l(l+1)$, one can write also 
\begin{eqnarray}\label{rkHy}
\left\langle \frac{1}{r} \right\rangle &=& \frac{\eta}{N_{Hy}^2}, \nonumber \\
\left\langle \frac{1}{r^2} \right\rangle &=& \frac{2 \eta^2}{(2 l+1)N_{Hy}^3}, \nonumber \\
\langle r \rangle   &=& \frac{1}{2 \eta} \left( 3 N_{Hy}^2 - L \right), \nonumber \\
\langle r^2 \rangle &=& \frac{N_{Hy}^2}{2 \eta^2} \left( 5 N_{Hy}^2 - 3 L + 1 \right), \nonumber \\
\langle r^3 \rangle &=& \frac{N_{Hy}^2}{8 \eta^3} \left( 35 N_{Hy}^4 + 5 N_{Hy}^2 (5 -6 L) + 3 L (L-2) \right), \nonumber \\
\langle r^4 \rangle &=& \frac{N_{Hy}^4}{8 \eta^4} \left( 63 N_{Hy}^4 + 35 N_{Hy}^2 (3 -2 L) + 5 L (3 L-10) +12 \right).
\end{eqnarray}
Using (\ref{p2b}) and (\ref{p4b}), it is easy to show that 
\begin{eqnarray}\label{pkHy}
\langle p^2 \rangle &=& \frac{\eta^2}{N_{Hy}^2}, \nonumber \\
\langle p^4 \rangle &=& \eta^4 \frac{8 n + 2 l+5}{(2 l+1)N_{Hy}^4}.
\end{eqnarray}

\section{Harmonic oscillator}\label{A_OH}

The eigenenergies of the harmonic oscillator Hamiltonian \cite{flu,brau99}
\begin{equation}\label{HOH}
H = \frac{\bm p^2}{2 m}+  \nu \, r^2
\end{equation}
are given by ($\hbar=1$)
\begin{equation}\label{EOH}
E = \sqrt{\frac{2 \nu}{m}} N_{HO} \quad \textrm{with} \quad N_{HO}=2 n +l+\frac{3}{2},
\end{equation}
and the normalized eigenvectors by
\begin{equation}\label{psiOH}
\psi_{nl\mu}(\bm r)=\lambda^{3/2}\sqrt{\frac{2 \; n!}{\Gamma(n+l+3/2)}}\left( \lambda r \right)^l e^{-\lambda^2 r^2/2}L_{n}^{l+1/2}(\lambda^2 r^2) Y^l_\mu(\hat{\bm{r}}),
\end{equation}
with $\lambda=\left( 2 m \nu \right)^{1/4}$. At the origin, the S-states are such that 
\begin{equation}\label{psi0OH}
|\psi_{n,0}(0)|^2=\lambda^3 \frac{2 \Gamma(n+3/2)}{\pi^2 n!}.
\end{equation}
Mean values $\langle r^k \rangle$ can be computed by performing directly the integrals or by using (\ref{vir5}). One obtains
\begin{equation}\label{rkHOgen}
\langle r^k \rangle = \frac{1}{\lambda^k}\frac{\Gamma(n+l+3/2)}{n!}\sum_{p,q=0}^n (-1)^{p+q}C_n^p C_n^q \frac{\Gamma(l+p+q+(k+3)/2)}{\Gamma(p+l+3/2) \Gamma(q+l+3/2)}. 
\end{equation}
One can also write
\begin{eqnarray}\label{rkOH}
\langle n|r |n\rangle &=& \frac{1}{\lambda} \frac{4 \Gamma(n+3/2)}{\pi n!}, \nonumber \\
\langle n,l|r^2 |n,l\rangle &=& \frac{N_{HO}}{\lambda^2}, \nonumber \\
\langle n|r^3 |n\rangle &=& \frac{1}{\lambda^3} \frac{8 (4 n+3) \Gamma(n+3/2)}{3 \pi n!}, \nonumber \\
\langle n,l|r^4 |n,l\rangle &=& \frac{1}{4 \lambda^4} \left( 6 N_{HO}^2 - 2 L + \frac{3}{2} \right).
\end{eqnarray}
Using the Fourier transform of a harmonic oscillator, it is easy to show that 
\begin{equation}\label{pkOH}
\langle p^k \rangle = \lambda^{2 k} \langle r^k \rangle .
\end{equation}

\section{Overlap of eigenfunctions}\label{A_overlap}

The scalar product of two radial functions $R_{n,l}(r)$ and $R_{n',l}(r)$ of a set of orthonormal states is simply given by $\delta_{nn'}$. When one of these functions is scaled by a positive factor $a$, the overlap 
\begin{equation}\label{ovgen}
F_{n,n',l}(a)=a^{3/2}\int_0^\infty R_{n,l}(x)R_{n',l}(a x)x^2 dx
\end{equation}
satisfies the following properties:
\begin{eqnarray}\label{ovprop}
&&\lim_{a\to 1} F_{n,n',l}(a)=\delta_{nn'}, \nonumber \\
&&|F_{n,n',l}(a)| \le 1, \nonumber \\ 
&&F_{n,n',l}(1/a)=F_{n',n,l}(a), \nonumber \\
&&\lim_{a\to 0} F_{n,n',l}(a) = \lim_{a\to \infty} F_{n,n',l}(a)=0.
\end{eqnarray}
The first relation stems from the definition~(\ref{ovgen}), the second one from the Schwarz inequality, and the others are due to scaling properties. 

Using the dilation properties of the Laguerre polynomials and the various existing recurrence relations \cite{grad80}, it is possible to compute analytically the formula $F_{n,n',l}(a)$ for hydrogen-like systems and harmonic oscillators. In the case of hydrogen-like systems, one obtains
\begin{eqnarray}\label{FHy}
F^{Hy}_{n,n',l}(a)&=&(-1)^{n+n'}\sqrt{a\, n!\, (N+l)!\, n'!\, (N'+l)!}\left( 4 a N N' \right)^N \frac{Q(a)^{n'-n}}{S(a)^{N'+N+1}} \nonumber \\
&&\times\sum_{k=0}^n (-1)^k \left(\frac{Q(a)^2}{4 a N N'}\right)^k \frac{1}{k!\, (n-k)!\, (N-k+l)!\, (n'-n+k+1)!} \nonumber \\
&&\times \left( 2(N-k)(n'-n+k+1) + (n-k)(N-k+l)\frac{Q(a)}{2 a N} + (n'-n+k)(n'-n+k+1)\frac{2 a N}{Q(a)} \right)
\end{eqnarray}
with $N=n+l+1$, $N'=n'+l+1$, $Q(a)=a N-N'$, and $S(a)=a N+N'$. In the case of harmonic oscillators, the formula is given by \cite{sema95}
\begin{eqnarray}\label{FOH}
F^{HO}_{n,n',l}(a)&=&\sqrt{n!\, n'!\, \Gamma(n+l+3/2)\, \Gamma(n'+l+3/2)} (2 a)^{2 n+l+3/2} \frac{\left( 1-a^2 \right)^{n'-n}}{\left( 1+a^2 \right)^{n+n'+l+3/2}} \nonumber \\
&&\times\sum_{k=0}^n (-1)^k \frac{\left( 1-a^2 \right)^{2 k}}{(2 a)^{2 k} k!\, (n-k)!\, (n'-n+k)!\, \Gamma(n-k+l+3/2)}.
\end{eqnarray}
It can be checked that both (\ref{FOH}) and (\ref{FHy}) satisfy relations~(\ref{ovprop}), but the calculations are much more tedious for $F^{Hy}$. 

\section{Solving a special equation}\label{A_invf}

The multivalued Lambert function $W(x)$ is defined by $W^{-1}(x)=x e^x$ which implies that $W(x) e^{W(x)}= W(x e^x)=x$ (see Ref.~\cite{af3} for more details). In section~\ref{exp}, it is necessary to solve the following equation (giving $x$ as function of the parameter $z$):
\begin{equation}\label{Wxal}
z = W(x) x^\alpha.
\end{equation}
Using the two changes of variables $x=y e^y$ and $\alpha y=(\alpha+1) u$, this equation can be inverted to give:
\begin{alignat}{2}\label{iWxal}
x&=W^{-1}(z)=z e^z &&\quad \textrm{if}\quad \alpha =0, \nonumber \\
x&=W^{-1}\left(\ln \frac{1}{z}\right)=\frac{1}{z}\ln \frac{1}{z} &&\quad\textrm{if} \quad  \alpha =-1, \nonumber \\
x&=W^{-1}\left(\frac{\alpha+1}{\alpha} W\left( \frac{\alpha}{\alpha+1} z^{1/(\alpha+1)}\right)\right) &&\quad\textrm{otherwise}.
\end{alignat}
The domains of the functions $z(x)$ and $x(z)$ must be examined carefully. They depends on the branch $W_0$ or $W_{-1}$ considered for the Lambert function and on the value of $\alpha$.

\end{document}